\documentclass[aps,a4paper,amsmath,amssymb,nofootinbib,floatfix,english]{revtex4-2}

\usepackage{physics, mathtools, babel, palatino, parskip, floatpag}
\usepackage[colorlinks=true]{hyperref}
\usepackage{graphicx}
\usepackage{subcaption}
\usepackage{pdfpages}
\usepackage{etoolbox}
\usepackage{caption}

\makeatletter
\patchcmd{\@outputpage@head}{\@ifx{\LS@rot\@undefined}{}{\LS@rot}}{}{}{}
\makeatother

\def\excerptA{\section*{Appendix A - Code for Raychaudhari Equation of Timelike Geodesics}}
\def\excerptB{\section*{Appendix B - Code for Raychaudhari Equation of Nulllike Geodesics}}

\date{\today}

\begin{document}
\title[Geodesic Congruences and the Raychaudhuri Equations in $\textrm{SAdS}_4$]{On Geodesic Congruences and the Raychaudhuri Equations in $\textrm{SAdS}_4$ Spacetime}
\author{D Biswas$^1$\footnote{Email: dripto.biswas@niser.ac.in} and J Shivottam$^1$\footnote{Email: jyotirmaya.shivottam@niser.ac.in\\}}
\address{$^1$ School of Physical Sciences, National Institute of Science Education and Research, Jatni Rd.,
Odisha - 752050, India.}

\begin{abstract}
In this article, we look into geodesics in the Schwarzschild-Anti-de Sitter metric in (3+1) spacetime dimensions. We investigate the class of marginally bound geodesics (timelike and null), while comparing their behavior with the normal Schwarzschild metric. Using \textit{Mathematica}, we calculate the shear and rotation tensors, along with other components of the Raychaudhuri equation in this metric and we argue that marginally bound timelike geodesics, in the equatorial plane, always have a turning point, while their null analogues have at least one family of geodesics that are unbound. We also present associated plots for the geodesics and geodesic congruences, in the equatorial plane.
\end{abstract}

\maketitle

\section{Introduction}\label{sec:intro}
The Schwarzschild solutions to the Einstein Field Equations (EFE) \cite{sch} have been studied in detail for a long time, being the simplest, non-trivial, spherically symmetric solution. The geodesic orbits (both timelike and null) have been investigated in various articles \cite{resca}. Another class of interesting spacetime solutions to the EFE are the constant curvature, maximally symmetric Anti-de Sitter (AdS) spacetime. The geodesics for AdS spacetime have been investigated in \cite{nyugen}. It was shown in the seminal work of Maldacena, that they correspond to strongly interacting field theories via the celebrated AdS/CFT correspondence principle \cite{maldacena,makoto}.

In this paper, we look into the Schwarzschild Anti-de Sitter ($\textrm{SAdS}_4$) solution in $(3+1)$ spacetime dimensions in Sec. \ref{sec:intro} and proceed to study its geodesic congruences (for timelike and null cases) in Secs. \ref{sec:timegeod} and \ref{sec:nullgeod}. The geodesic equations for this case have been studied in detail in \cite{hackmann}. We also describe the varied behaviours of the congruences based on the initial parameters of the metric and confirm the same by simulating them in $Mathematica$. Furthermore, we calculate the congruence evolution, using the Raychaudhuri equations (see \cite{sayan} for a detailed discussion) for timelike and null cases and derive the geometric properties (shear and rotation tensors) for a certain class of geodesics, namely marginally bound geodesics, in $\textrm{SAdS}_4$.

\section{The $\textrm{SAdS}_4$ metric and marginally bound geodesics}\label{sec:metric}
The $\textrm{SAdS}_4$ line element in Schwarzschild coordinates is given by,
\begin{equation}\label{metric}
       \dd s^2 = -f \dd t^2 + f^{-1}\dd r^2 +r^2 \dd \theta^2 + r^2\sin^2\theta \dd \phi^2,
\end{equation}
where $f = (1 - 2M/r + r^2/a^2)$. Here, $M$ denotes the mass of the Schwarzschild black hole (SBH) and $a$ denotes the AdS radius. 
The radius of the $\textrm{SAdS}$ BH ($r_s$) satisfies,
\begin{equation}\label{rh}
r_s^3 -2Ma^2 + a^2 r_s =0,
\end{equation}
which has been shown to have only one real root \cite{soco}.
One may immediately note two Killing vectors associated with Eq. (\ref{metric}), given by,
$K^{\nu}_{(t)} = (1,0,0,0)$ and $K^\nu_{(\phi)} = (0,0,0,1)$. To simplify our calculations henceforth, we shall focus on marginally bound geodesics, which satisfy, $u_t = -1$. Most of the interesting properties of geodesic congruences are also obtained in this case as we shall see in the following sections. In the rest of the article, we work in the equatorial plane $\theta = \pi/2$ to further simplify calculations \footnote{It is easy to check that, for a particle/light ray, initially in the equatorial plane, its motion/path is always confined to this plane.}.

\section{Timelike Geodesic Congruences}\label{sec:timegeod}
Timelike radial geodesic congruences in the Schwarzschild metric have been explained in the book by Poisson (see \cite{poisson} ch. 2). We shall see that by taking the proper limits of our results in this section, these well-established results are recovered. For a timelike geodesic, we use the proper time as the affine parameter, i.e. $u^\alpha = \dd{x^\alpha}/\dd{\tau}$. Then, we have, from Sec. \ref{sec:metric},
\begin{equation}\label{ucontra}
u^t = g^{tt} u_t = f^{-1}.
\end{equation}
Using the Killing vector $K^\nu_{(\phi)} = (0,0,0,1)$ for the metric in Eq. (\ref{metric}), we observe that the quantity,
\begin{equation}\label{angMom}
L = m g_{\mu\nu}K^\nu_{(\phi)}u^\mu = mr^2u^\phi ,
\end{equation}
is conserved. Here, $L$ is the orbital angular momentum, and $m$ denotes the mass of the test particle following the geodesic. We use the normalization condition, $g_{\mu\nu}u^\mu u^\nu = -1$ for timelike geodesics, to obtain,
\begin{equation}\label{norm}
    g_{tt}(u^t)^2 + g_{rr}(u^r)^2 + g_{\phi\phi}(u^\phi)^2 = -\frac{1}{f} + \frac{1}{f}(u^r)^2 + r^2\left(\frac{\mathcal{L}_T}{r^2}\right)^2 = -1,
\end{equation}
where $\mathcal{L}_T = L/m$. Simplifying Eq. (\ref{norm}), we get,
\begin{equation}\label{urR}
u^r = \pm\sqrt{1 - f - \frac{\mathcal{L}_T^2}{r^2}f} = \pm\sqrt{\frac{2M}{r}-\frac{r^2}{a^2} - \frac{\mathcal{L}_T^2}{r^2}\left(1-\frac{2M}{r}+\frac{r^2}{a^2}\right)},
\end{equation}
where the positive sign is for outgoing geodesics.

\subsection{Calculating the expansion}\label{subsec:THT}
The expansion $\Theta_T$ for geodesics with tangent parameter, $u_\alpha$, is simply given by,
\begin{equation}\label{ThT}
    \Theta_T = u^\alpha_{;\alpha} = \frac{1}{r^2}(r^2 u^r),r = \pm\frac{a^2(\mathcal{L}_T^2(M-r)+3Mr^2)-2\mathcal{L}_T^2r^3-3r^5}{a^2r^4\sqrt{\frac{\mathcal{L}_T^2(2M-r)}{r^3}+\frac{2M}{r}-\frac{\mathcal{L}_T^2+r^2}{a^2}}}.
\end{equation}
One can note that,
\begin{equation}\label{limitThT}
    \lim_{a \rightarrow \infty} \Theta_T |_{\mathcal{L}_T=0} = \pm\frac{3}{2}\sqrt{\frac{2M}{r^3}}.
\end{equation}

Eq. (\ref{limitThT}) matches exactly with the result for radial, marginally bound timelike geodesics (MTGC) of the Schwarzschild metric, shown in Poisson \cite{poisson} (Sec. 2.3.7).

Moreover, one can show that the rotation tensor $\omega_{\alpha\beta} = u_{\alpha;\beta} - u_{\beta;\alpha} = 0$ for MTGC (refer to Appendix A). This implies, that these MTGC are \textit{hypersurface orthogonal}, by the Frobenius Theorem \cite{poisson,wald}.

\subsection{Timelike geodesic simulations and the different types of congruence expansion
plots}\label{subsec:timesimul}
In all the relevant figures of this article, the BH horizon ($r_s$) is depicted by an orange circle for geodesic plots and by a vertical orange line for $\Theta$ vs $r$ plots. Similarly, all congruence singularities are denoted by dotted black lines (or circles).

The non-zero Christoffel symbols for the $\textrm{SAdS}_4$ metric are provided in Appendix A. Here, we start by writing out the radial geodesic equation for MTGC, which simplifies to:
\begin{equation}
    \dv{u^r}{\tau} + \frac{M}{r^2} + \frac{r}{a^2} - \frac{L^2 (-3 M + r)}{r^4} = 0
    \label{geodRadTime}.
\end{equation}

\begin{figure}
        \centering
        \begin{subfigure}[b]{0.5\textwidth}
                \centering
                \includegraphics[width=0.82\textwidth]{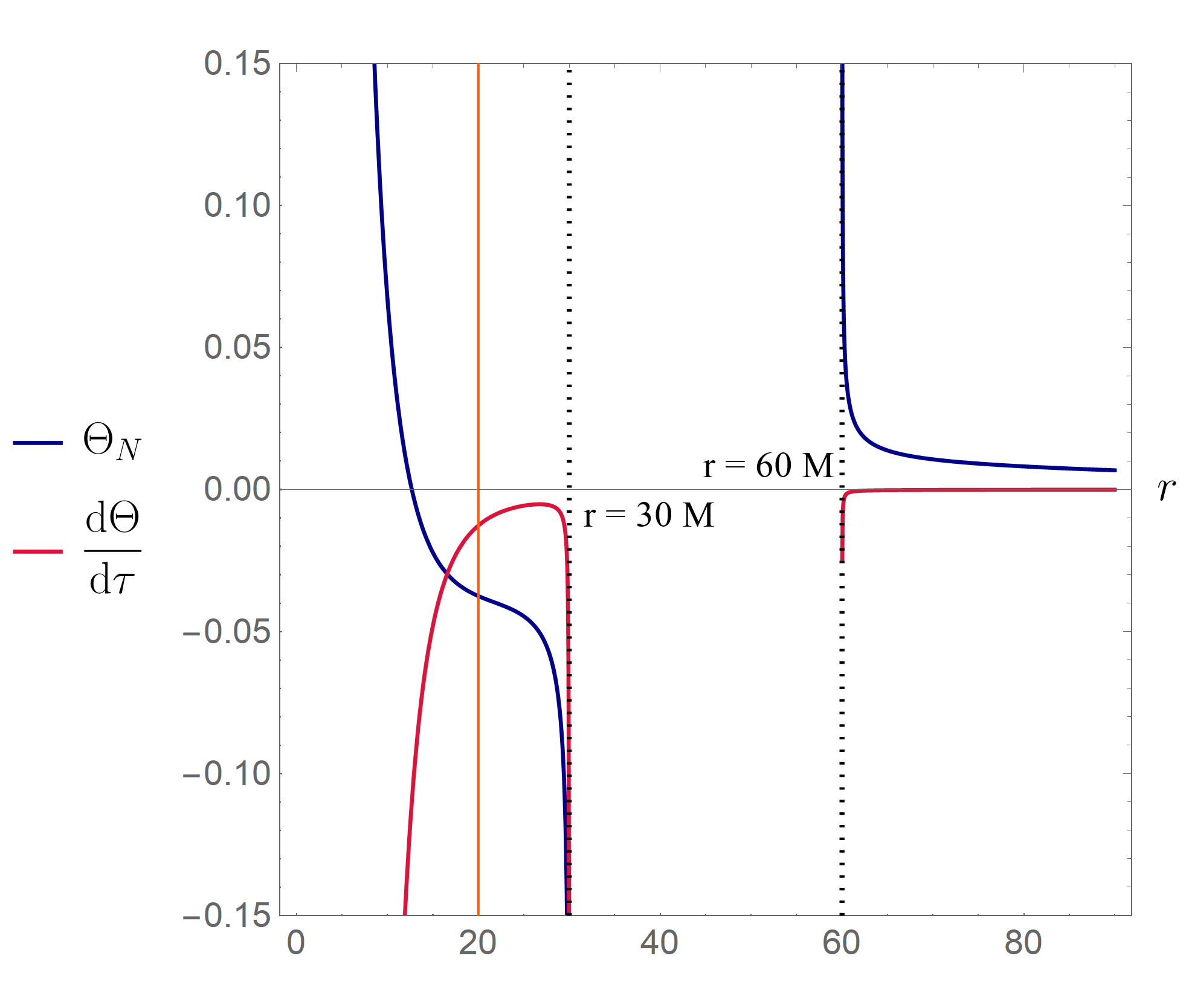}
                \caption{MTGC, Region 1}
                \label{fig:gulltime1}
        \end{subfigure}%
        \begin{subfigure}[b]{0.5\textwidth}
                \centering
                \includegraphics[width=0.82\textwidth]{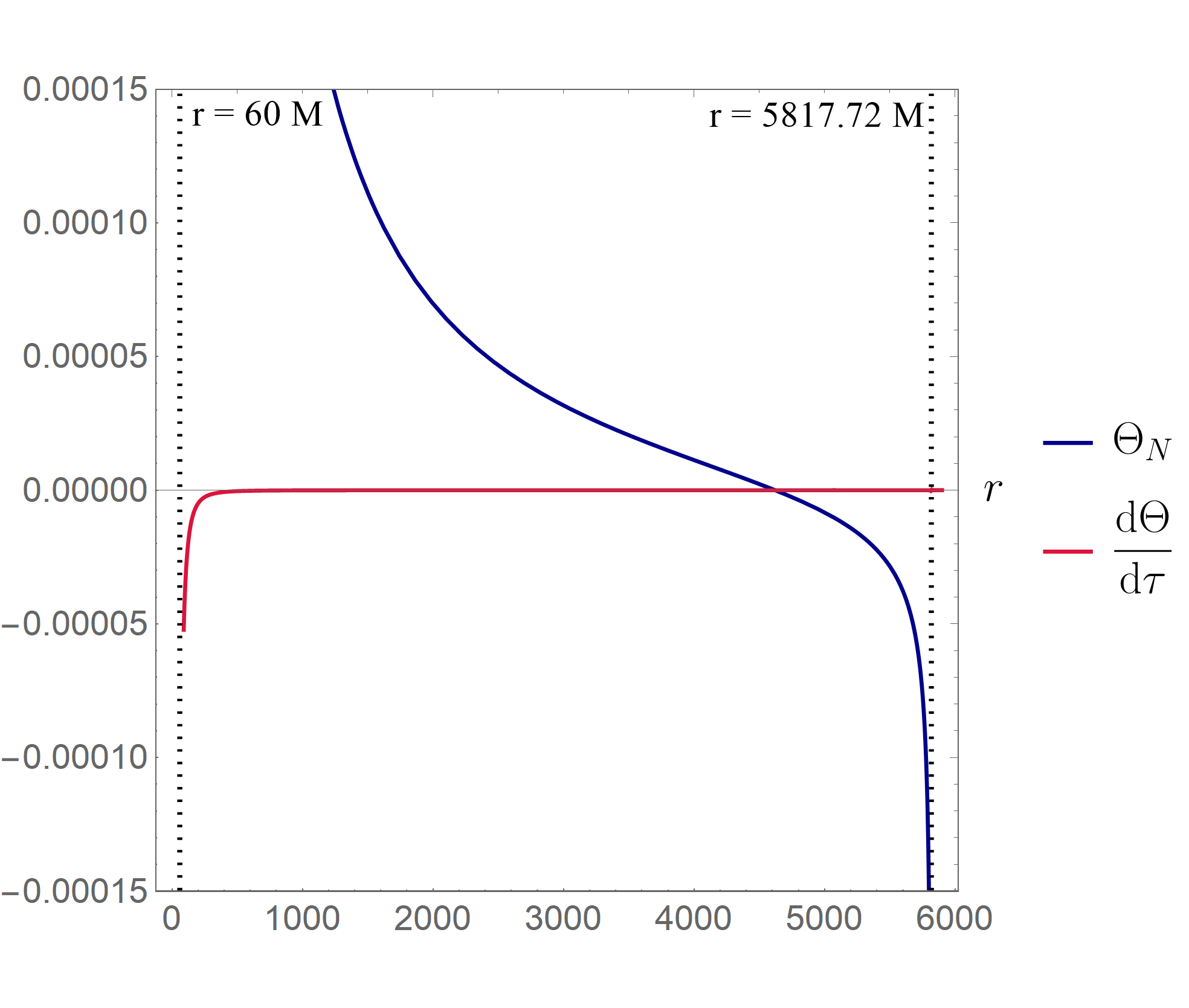}
                \caption{MTGC, Region 2}
                \label{fig:gulltime2}
        \end{subfigure}%

        \begin{subfigure}[b]{0.5\textwidth}
                \centering
                \includegraphics[width=\textwidth]{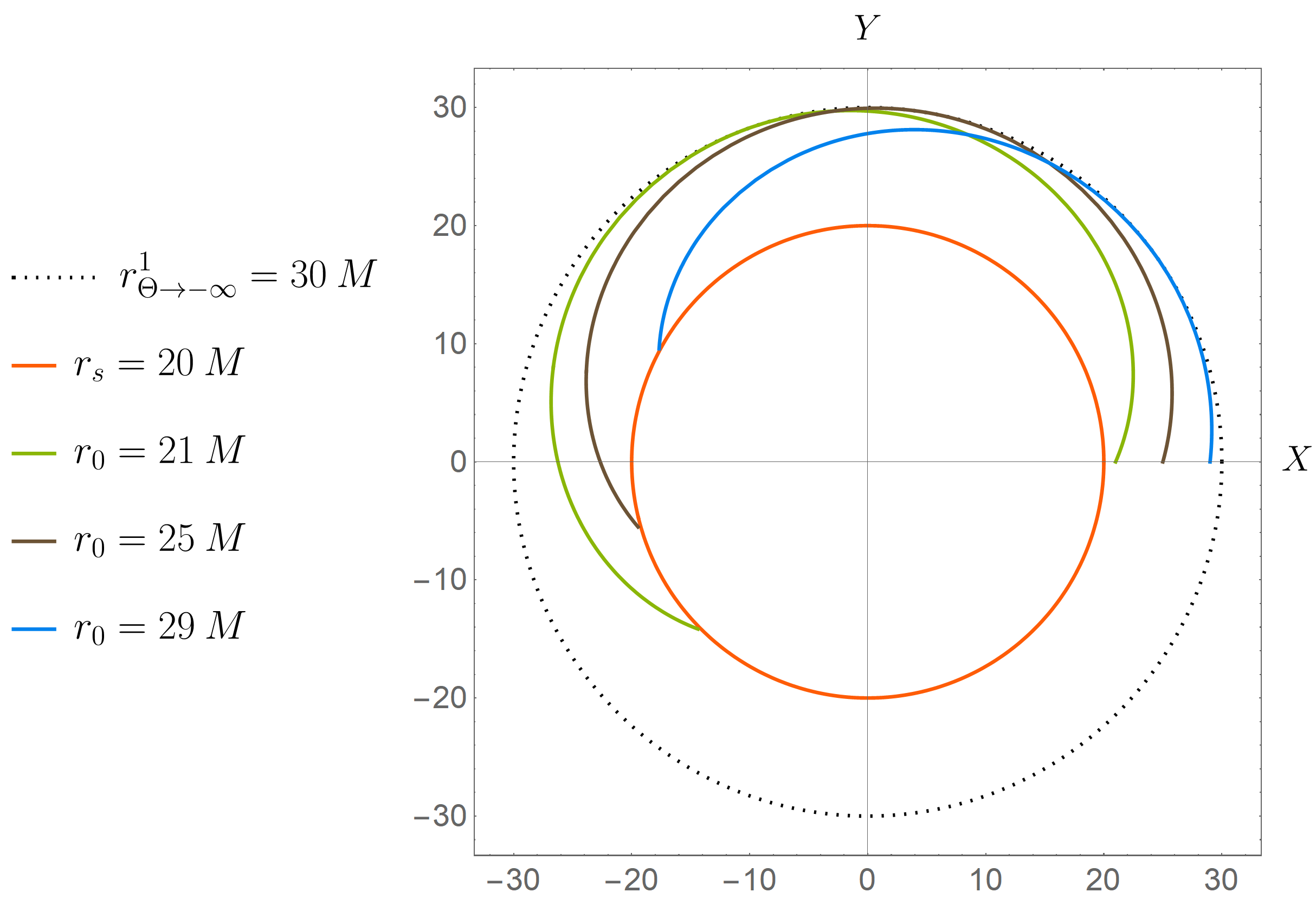}
                \caption{Bounded Timelike Geodesics, Region 1}
                \label{fig:timebound1}
        \end{subfigure}%
        \begin{subfigure}[b]{0.5\textwidth}
                \centering
                \includegraphics[width=\textwidth]{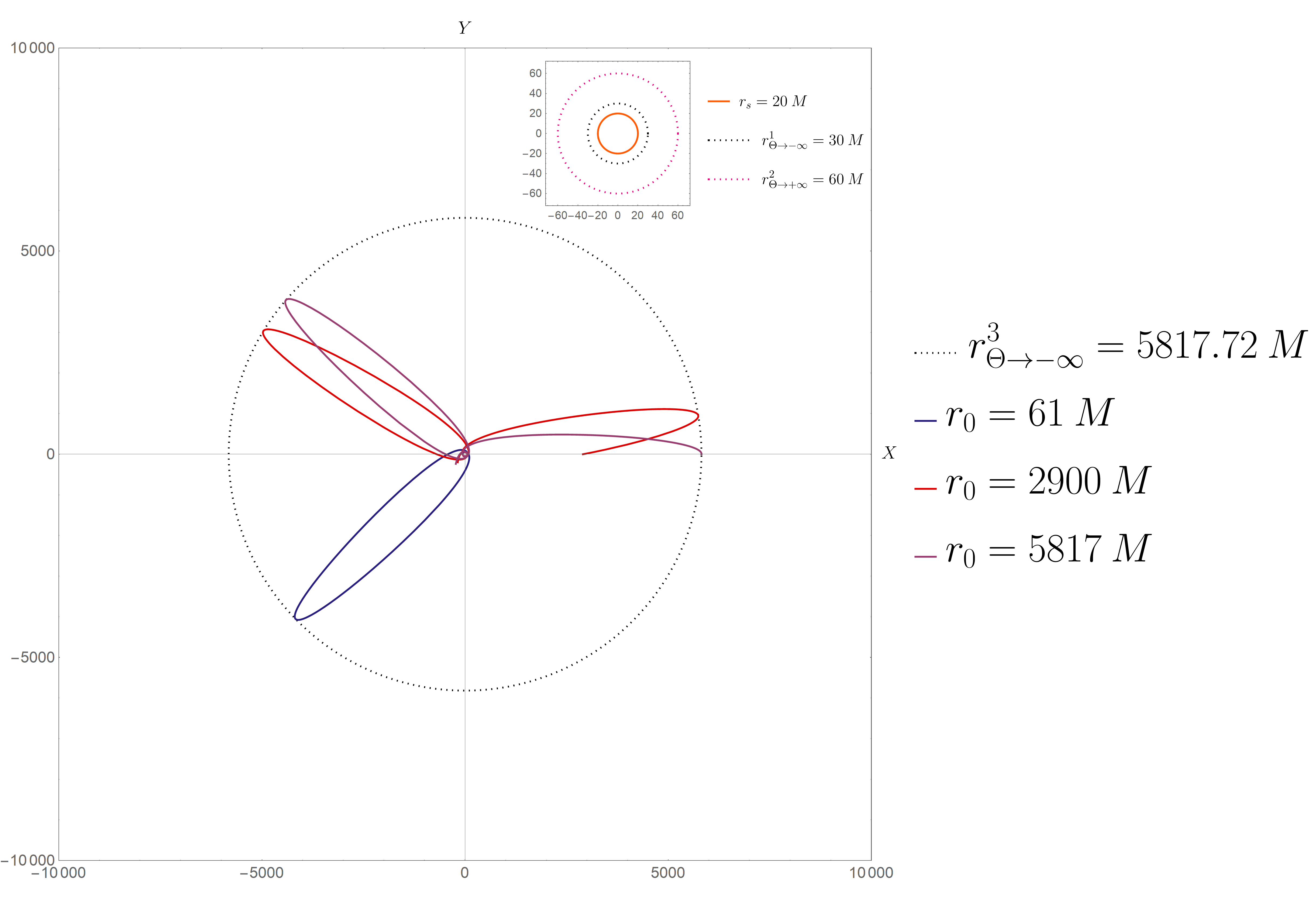}
                \caption{Bounded Timelike Geodesics, Region 2}
                \label{fig:timebound2}
        \end{subfigure}
        \caption{(a) Example of an MTGC with two $\textit{bound}$ branches. (b) The second \textit{bound} branch at large $r$. (c) A family of \textit{bound} timelike geodesics starting at $r_0 < 30$ M, corresponding to Region 1. (d) A similar \textit{bound} family corresponding to Region 2.}\label{fig:timecases}
\end{figure}

Using Eq. (\ref{geodRadTime}), we obtained two broad classes of geodesic congruences. In this article, we call a congruence, $\textit{bound}$, if the radial parameter ($r$) of the family of geodesics is bounded by a maximal finite value, i.e. the geodesics have a turning point. Typical plots of bound congruences and their evolution are shown in Fig. \ref{fig:timebound1}, \ref{fig:timebound2} and Fig. \ref{fig:nullbound}. On the other hand, a congruence, whose radial parameter is unbounded, is referred to as \textit{asymptotic}. An example has been depicted in Fig. \ref{fig:nullasymp}. Refer to the respective sections, along with Sec. \ref{subsec:largeR} for further details and discussion.
It can be noticed that in Fig. \ref{fig:timebound2}, the elliptical orbits precess over large proper times and have a perihelion radius $r_p \ge 60$ M. This is in accordance with the $+\infty$ singularity of the second branch of the congruence in Fig. \ref{fig:gulltime1} , \ref{fig:gulltime2} at $r^1_{\Theta\rightarrow-\infty}$.

\section{Null Geodesic Congruences}\label{sec:nullgeod}
Taking $k^\alpha = \dd{x^\alpha}/\dd{\lambda}$, where $\lambda$ is some affine parameter (but not proper time), we proceed in a similar fashion to Sec. \ref{sec:timegeod}. Then, we have, $k_t = -1$, since we are considering marginally bound, null geodesic congruences (MNGC), and $k^t = f^{-1}$. Also, we have,
\begin{equation}\label{AngMomNull}
\mathcal{L}_N = g_{\mu\nu}K^\nu_{(\phi)}k^\mu = r^2k^\phi,
\end{equation}
where $\mathcal{L}_N$ is some conserved angular momentum-like quantity similar to $\mathcal{L}_T$ above. We shall see from the simulated plots, that $\mathcal{L}_N$ indeed governs, how much the MNGC curve around the black hole. From the normalization condition, we get,
\begin{equation}\label{urNull}
    k^r = \pm\sqrt{1 - \frac{\mathcal{L}_N^2}{r^2}\left(1 - \frac{2M}{r} + \frac{r^2}{a^2}\right)}.
\end{equation}

\subsection{Calculating the expansion}\label{subsec:THN}
Starting from the general expression for congruence expansions, we get,
\begin{equation}\label{ThN}
\Theta_N = k^\alpha_{;\alpha} = \pm\frac{1}{r^2}(r^2 k^r)_{,r} = \frac{2 a^2 r^3 + \mathcal{L}_N^2 [a^2 (M - r) - 2 r^3]}{a^2 r^4 \sqrt{1 - \frac{\mathcal{L}_N^2}{a^2 r^3}\left(r^3 + a^2 (-2 M + r)\right)}}.
\end{equation}
Again, we see that,
\begin{equation}\label{nulllim}
\lim_{a\rightarrow \infty} \Theta_N|_{\mathcal{L}_N=0} = \pm \frac{2}{r},
\end{equation}
which is the expression obtained in Sec. 2.4.7 of Poisson \cite{poisson}, for radial null geodesics in normal Schwarzschild spacetime. Furthermore, we have calculated the shear and rotation tensors and showed that $\omega_{\alpha\beta} = 0$ for $\theta = \pi/2$ (refer to Appendix B).
\subsection{Null geodesic simulations and different types of expansion plots}\label{subsec:nullsimul}
\begin{figure}
        \centering
        \begin{subfigure}[b]{\textwidth}
                \centering
                \includegraphics[width=0.4\textwidth]{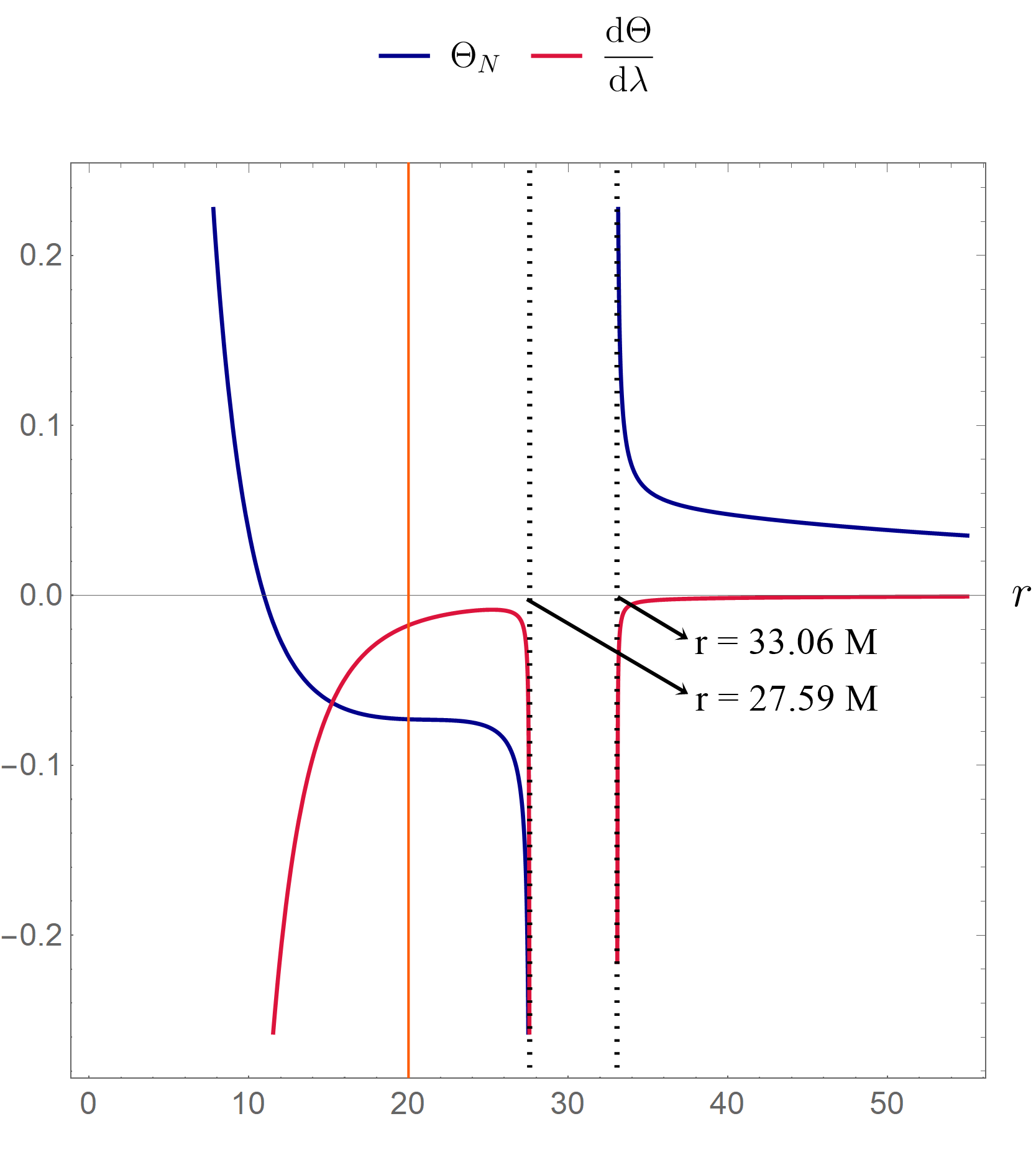}
                \caption{MNGC}
                \label{fig:nullcong1}
        \end{subfigure}%

        \begin{subfigure}[b]{0.5\textwidth}
                \centering
                \includegraphics[width=\textwidth]{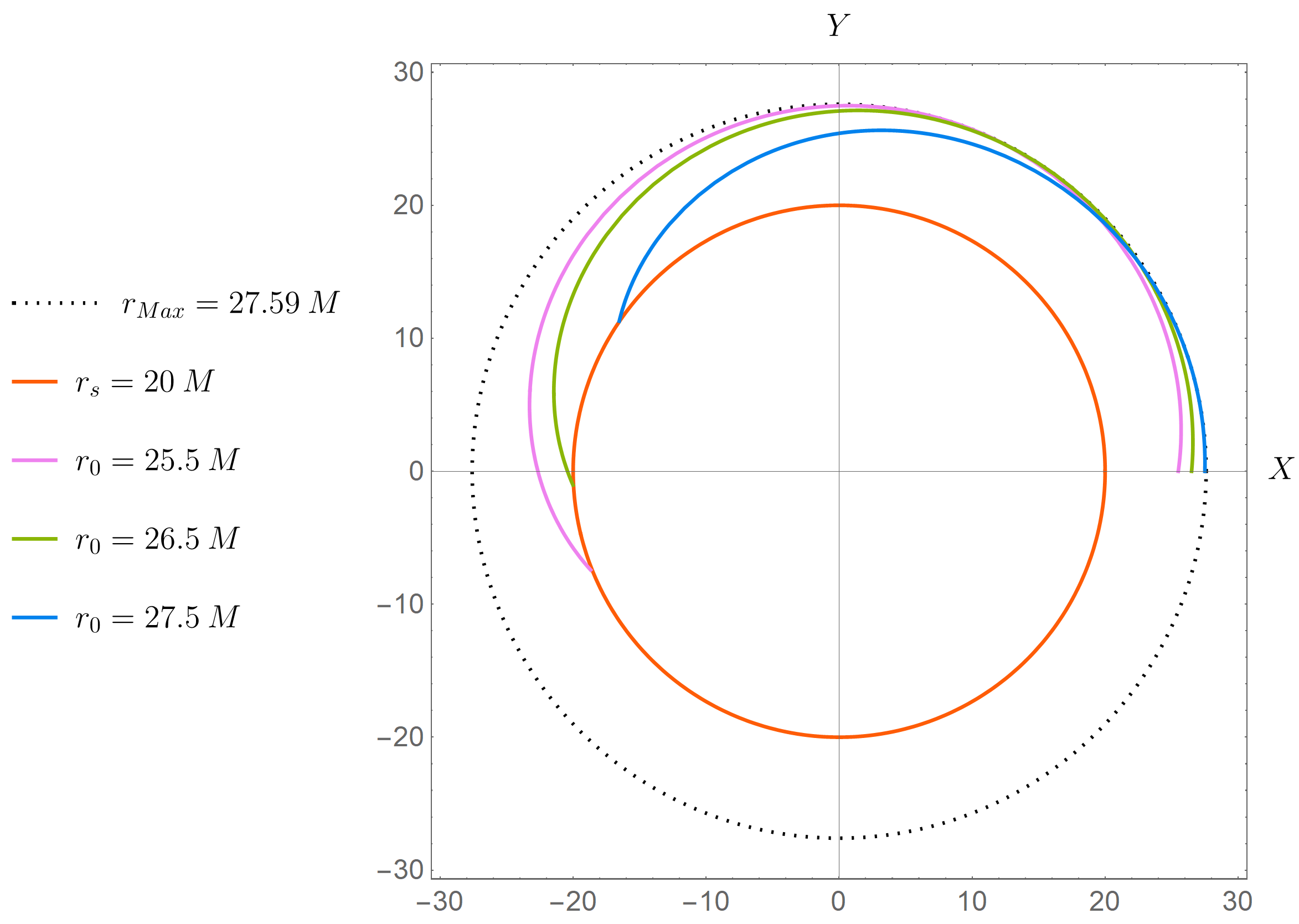}
                \caption{Bounded Null Geodesics}
                \label{fig:nullbound}
        \end{subfigure}%
        \begin{subfigure}[b]{0.5\textwidth}
                \centering
                \includegraphics[width=\textwidth]{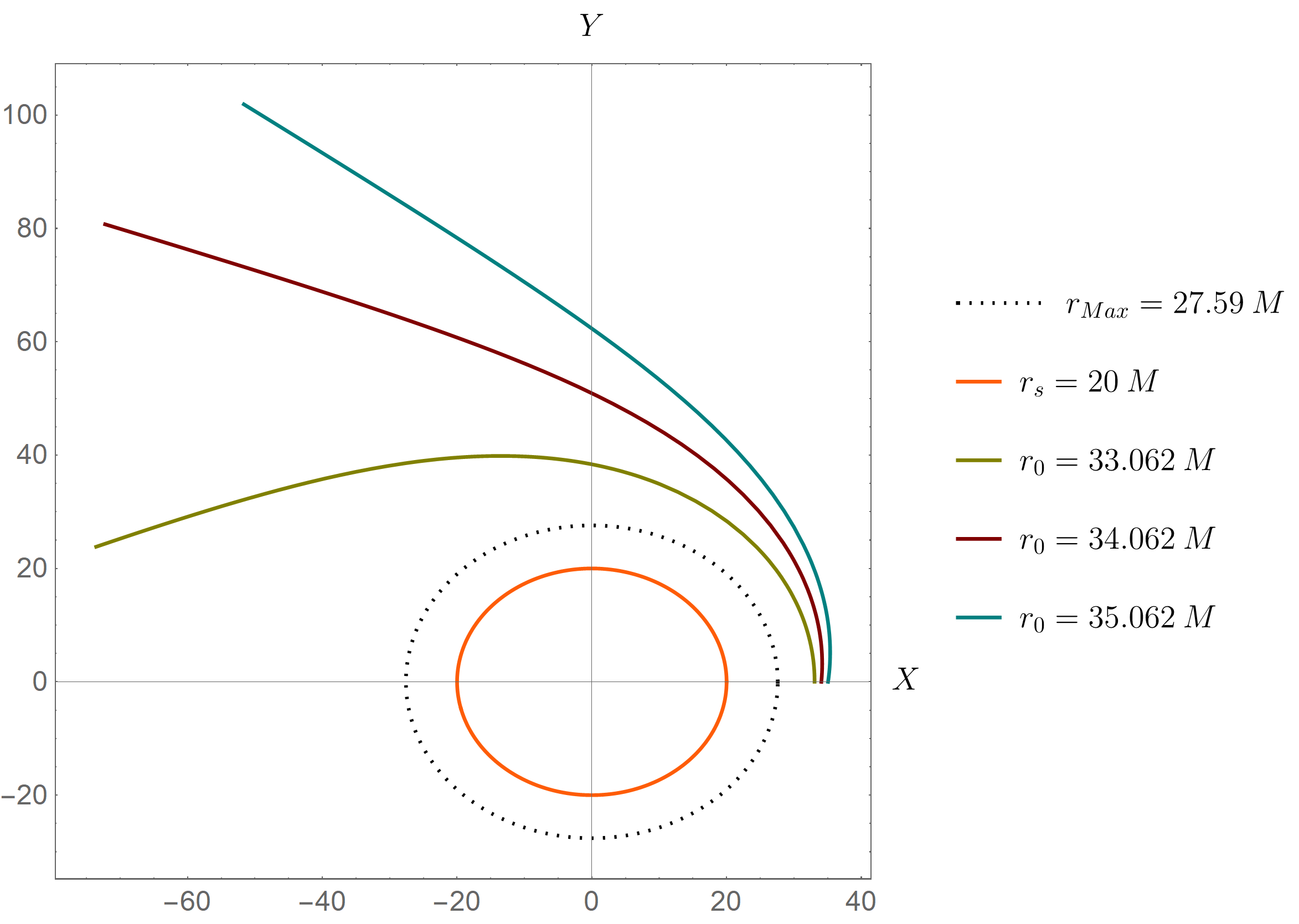}
                \caption{Asymptotic Null Geodesics}
                \label{fig:nullasymp}
        \end{subfigure}
        \caption{(a) Example of an MNGC with both $\textit{bound}$ and $\textit{asymptotic}$ branches. (b) A family of \textit{bound} null geodesics starting at $r_0 < 27.59$ M. (c) A family of \textit{asymptotic} null geodesics starting at $r_0 > 33.061$ M.}\label{fig:nullcases}
\end{figure}
Following the same procedure as Sec. \ref{subsec:timesimul}, we write out the simplified geodesic equation for the null case, which is given by,
\begin{equation}
    \dv{k^r}{\lambda} + \frac{\mathcal{L}_N^2 (r-3M)}{r^4} = 0.
    \label{geodRadNull}
\end{equation}
In Fig. \ref{fig:nullcases}, we have used Eq. (\ref{geodRadNull}) to simulate null geodesics, with various initial positions ($r_0$), for the parameters $(a = 10^6, M = 10, \mathcal{L}_N = 52.599999)$, in $\textit{Mathematica}$. As shown in Fig. \ref{fig:nullcong1}, the congruence, $\Theta_N$, for this case contains both \textit{bound} and \textit{asymptotic} branches. The corresponding plots are shown in Fig. \ref{fig:nullbound} and \ref{fig:nullasymp}. Note that, each of the lower plots in Fig. \ref{fig:nullcases} respects the corresponding congruence singularity.

\section{Raychaudhuri equations in $\textrm{SAdS}_4$}\label{sec:raychaudhuri}
Let us define $B_{\alpha\beta} = u_{\alpha;\beta}$. Then,
\begin{equation}\label{shear}
\sigma_{\alpha\beta} = B_{(\alpha\beta)} - \frac{1}{3}\Theta h_{\alpha\beta},
\end{equation}
is referred to as the shear tensor and,
\begin{equation}\label{rot}
\omega_{\alpha\beta} = B_{[\alpha\beta]},
\end{equation}
is the rotation tensor (which is zero in case of MNGC and MTGC, in the equatorial plane).
Then the general form of the Raychaudhuri equation for timelike geodesic congruences, is given by \cite{sayan,poisson},
\begin{equation}\label{RayT}
    \dv{\Theta_T}{\tau} = -\frac{1}{3}\Theta_T^2 - \sigma^{\alpha\beta}\sigma_{\alpha\beta} + \omega^{\alpha\beta}\omega_{\alpha\beta}-R_{\alpha\beta}u^\alpha u^\beta,
\end{equation}
where $R_{\alpha\beta}$ is the Ricci curvature tensor. We have computed the terms in Eq. (\ref{RayT}) in \textit{Mathematica}. The corresponding code and results have been included in Appendix A.

For the null case however, we need to find a transverse tensor, $\tilde{B}_{\alpha\beta}$. Following the discussion in Sec. 2.4.1 of Poisson \cite{poisson}, we choose an auxiliary null field, $n_\alpha$, and define $h_{\alpha\beta} = g_{\alpha\beta} + k_\alpha n_\beta + k_\beta n_\alpha$. Then, setting $\tilde{B}_{\alpha\beta} = h^\mu_\alpha h^\nu_\beta B_{\mu\nu}$, we can calculate $\sigma_{\alpha\beta}$ and $\omega_{\alpha\beta}$ using,
\begin{equation}\label{shearNull}
    \sigma_{\alpha\beta} = \tilde{B}_{(\alpha\beta)} - \frac{1}{2}\Theta h_{\alpha\beta},
\end{equation}
and Eq. (\ref{rot}) respectively (after replacing $B$ with $\tilde{B}$). For detailed expressions and calculations, pertinent to this section, the reader is pointed to the $\textit{Mathematica}$ code in Appendix B. The Raychaudhuri equations for the null case is then given by \cite{poisson},
\begin{equation}\label{RayNull}
    \dv{\Theta_N}{\lambda} = -\frac{1}{2}\Theta_N^2 - \sigma^{\alpha\beta}\sigma_{\alpha\beta} + \omega^{\alpha\beta}\omega_{\alpha\beta}-R_{\alpha\beta}k^\alpha k^\beta .
\end{equation}
Similar to the timelike case, one can check (from the code in Appendix B) that the quadratic invariant $\omega=\omega^{\alpha\beta}\omega_{\alpha\beta} = 0$ for $\theta=\pi/2$, i.e. $\omega_{\alpha\beta} = 0$ in the equatorial plane.

\subsection{Large $r$ behaviour of the congruences, $\Theta_T$ and $\Theta_N$}\label{subsec:largeR}
In this subsection, we shall see, that the large $r$ behaviour of $\Theta_T$ and $\Theta_N$ reveals some interesting properties of MTGC and MNGC respectively, which do not have any counterpart in the normal Schwarzschild or $\textrm{AdS}$ spacetimes. Quite simply, we observe that,
\begin{equation}\label{largeRTime}
\lim_{r\rightarrow \infty}\Theta_T = 3\sqrt{-\frac{1}{a^2}}.
\end{equation}
For any finite $a$, Eq. (\ref{largeRTime}) implies that all MTGC are bound, since at large $r$, $\Theta_T$ becomes complex, with no physical significance. Such a transition can only occur after the congruence singularity ($-\infty$) has been reached, because at small $r$, the denominator of the R.H.S of Eq. (\ref{ThT}) is positive. 

Similarly, we note that,
\begin{equation}\label{largeRNull}
\lim_{r\rightarrow \infty} \Theta_N = 0.
\end{equation}
We interpret this as all MNGC having at least one \textit{asymptotic} branch. Fig. \ref{fig:nullcong1} shows an example with both types of branches. It is also possible to have no bound branches for MNGC (for example, see Fig. \ref{fig:specialNull}).
\subsection{Plots for the congruences and their evolution}\label{subsec:THplots}
\begin{figure}
        \centering
        \begin{subfigure}[b]{0.5\textwidth}
                \centering
                \includegraphics[width=\textwidth]{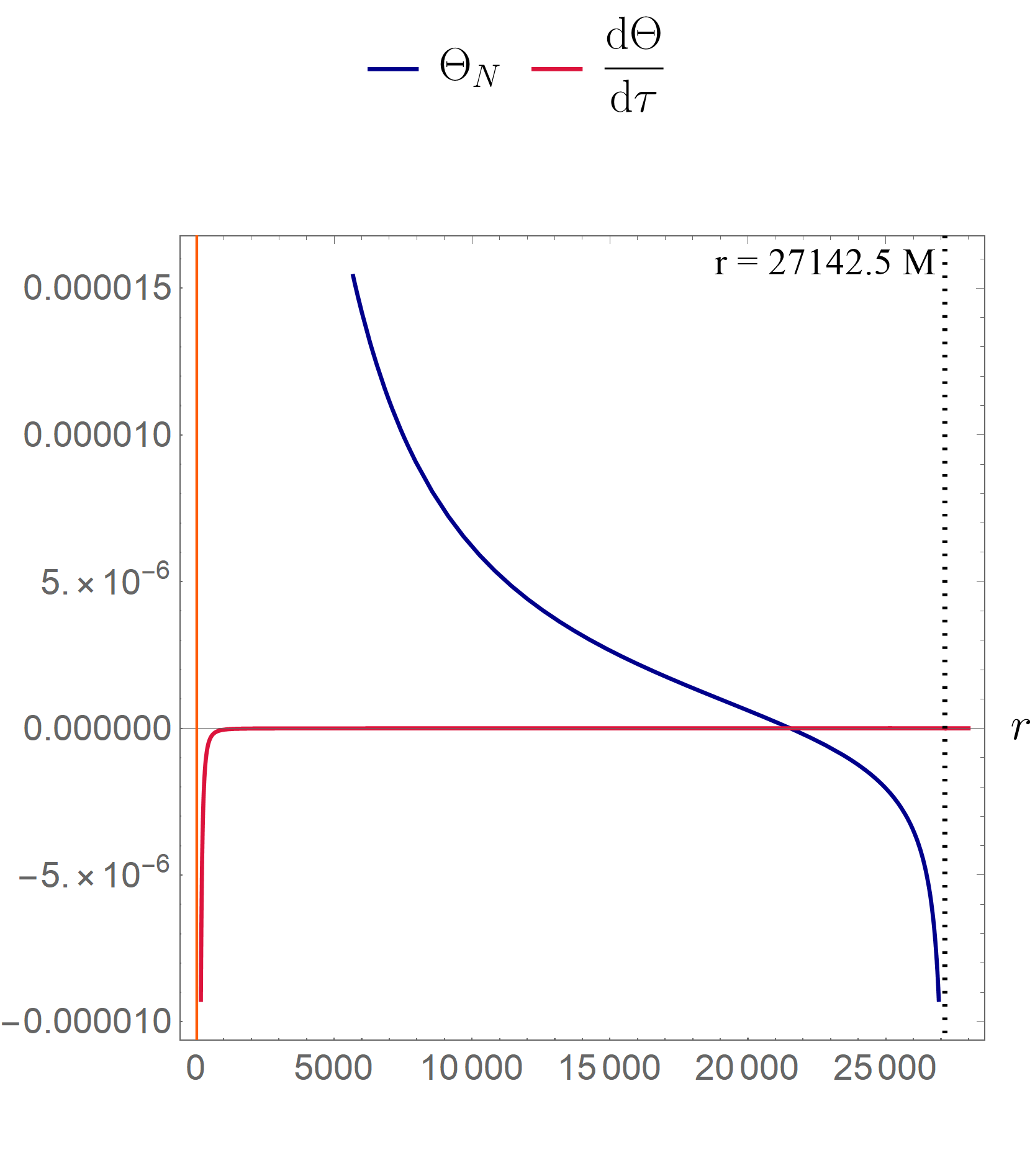}
                \caption{}
                \label{fig:TimeCong2}
        \end{subfigure}%
        \begin{subfigure}[b]{0.5\textwidth}
                \centering
                \includegraphics[width=\textwidth]{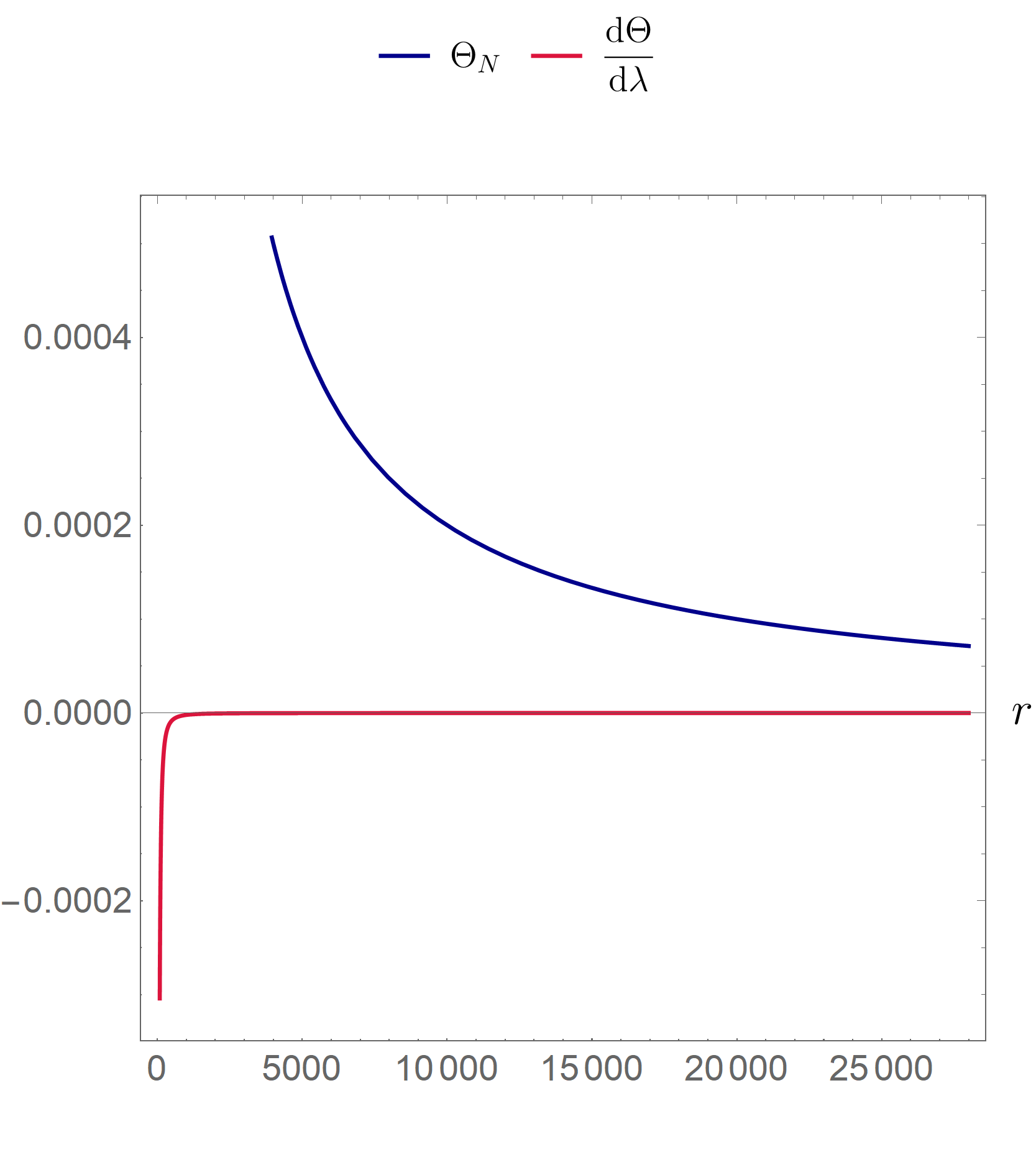}
                \caption{}
    \label{fig:specialNull}
        \end{subfigure}%
        \caption{(a) Plot for a \textit{bound} MTGC with parameters $(a = 10^6, M= 10, L = 10)$. $\dd{\Theta}/\dd{\tau}$ is negative throughout. (b) Plot for a purely \textit{asymptotic} congruence $\Theta_N$ and its evolution $\dd{\Theta_N}/\dd{\lambda}$ for parameters $(a = 10^6, M= 10, L = 10)$.}\label{fig:randomcases}
\end{figure}

Fig. \ref{fig:TimeCong2} is a typical example for a MTGC evolution.
It must be noted, that $\Theta_T$ does not extend beyond the congruence singularity (Fig. \ref{fig:TimeCong2}). 
This is because, $\Theta_T$ becomes complex after the congruence singularity at $r^0_{\Theta_T\rightarrow-\infty}=27142.5$ M.
MNGC, however, can be purely \textit{asymptotic}. In Fig. \ref{fig:specialNull}, we plot such an example of $\Theta_N$ and its evolution, $\dd{\Theta}/\dd{\lambda}$, with respect to $r$.

\section{Conclusion}\label{sec:summary}
In this article, we studied the expansions and their evolution, for MTGC and MNGC in $\textrm{SAdS}_4$ spacetime. The results have been occasionally contrasted with the normal Schwarzschild metric to point out important differences. We have also showed how the various properties reduce to the well-known results on taking proper limits. 

In Sec. \ref{sec:raychaudhuri}, we have used Eq. (\ref{RayT}) and (\ref{RayNull}) to calculate the congruence evolution. The detailed results may be found in the $\textit{Mathematica}$ code in Appendix A and B. The various geometric parameters, like shear and rotation, are also calculated separately for the MTGC and MNGC.

Finally, we note, that MNGC and MTGC exhibit significantly different behavior, as discussed in detail in the previous section. These are much more exotic, when compared to their counterparts in normal Schwarzschild metric. Moreover, as discussed in Sec. \ref{subsec:largeR}, for $\textrm{SAdS}_4$, we can only find \textit{bound} MTGC, while we can always find MNGC with both, \textit{bound} and \textit{asymptotic} branches, for all triples $(a,\mathcal{L},M)$. To the best of our knowledge, this behavior is unique to this metric (owing to the AdS properties via parameter, $a$, of this spacetime) and can have deep implications for further analysis. In the future, we hope to extend the study of Hawking-Penrose singularity theorems \cite{hawking} to $\textrm{SAdS}_4$ spacetime. Such a venture would require intricate details of the Raychaudhuri equations, some of which have been discussed in this paper.

\bibliographystyle{unsrt}
\bibliography{refs} % Entries

\newpage


\section*{Supplementary material}

# APPENDIX A - CODE FOR RAYCHAUDHARI EQUATION OF TIMELIKE GEODESICS

```mathematica
In[ ]:= << xAct`xTensor`
```

------------------------------------------------------------

Package xAct`xPerm`  version 1.2.3, {2015, 8, 23}

CopyRight (C) 2003-2020, Jose M. Martin-Garcia, under the General Public License.

Connecting to external MinGW executable...

Connection established.

------------------------------------------------------------

Package xAct`xTensor`  version 1.1.4, {2020, 2, 16}

CopyRight (C) 2002-2020, Jose M. Martin-Garcia, under the General Public License.

------------------------------------------------------------

These packages come with ABSOLUTELY NO WARRANTY; for details type
  Disclaimer[]. This is free software, and you are welcome to redistribute
  it under certain conditions. See the General Public License for details.

------------------------------------------------------------

```mathematica
In[ ]:= DefManifold[M4, 4, {α, β, γ, μ, ν, λ, σ, η}]
```

** DefManifold: Defining manifold M4.

** DefVBundle: Defining vbundle TangentM4.

```mathematica
In[ ]:= DefMetric[-1, metric[-α, -β], CD, {";", "∇"}, PrintAs → "g"]
```

** DefTensor: Defining symmetric metric tensor metric[-α, -β].

** DefTensor: Defining antisymmetric tensor epsilonmetric[-α, -β, -γ, -η].

** DefTensor: Defining tetrametric Tetrametric[-α, -β, -γ, -η].

** DefTensor: Defining tetrametric Tetrametric†[-α, -β, -γ, -η].

** DefCovD: Defining covariant derivative CD[-α].

** DefTensor: Defining vanishing torsion tensor TorsionCD[α, -β, -γ].

** DefTensor: Defining symmetric Christoffel tensor ChristoffelCD[α, -β, -γ].

** DefTensor: Defining Riemann tensor RiemannCD[-α, -β, -γ, -η].

** DefTensor: Defining symmetric Ricci tensor RicciCD[-α, -β].

** DefCovD:  Contractions of Riemann automatically replaced by Ricci.

** DefTensor: Defining Ricci scalar RicciScalarCD[].

** DefCovD:  Contractions of Ricci automatically replaced by RicciScalar.

** DefTensor: Defining symmetric Einstein tensor EinsteinCD[-α, -β].

** DefTensor: Defining Weyl tensor WeylCD[-α, -β, -γ, -η].

** DefTensor: Defining symmetric TFRicci tensor TFRicciCD[-α, -β].

** DefTensor: Defining Kretschmann scalar KretschmannCD[].

** DefCovD:  Computing RiemannToWeylRules for dim 4

** DefCovD:  Computing RicciToTFRicci for dim 4

** DefCovD:  Computing RicciToEinsteinRules for dim 4

** DefTensor: Defining weight +2 density Detmetric[]. Determinant.

```mathematica
In[ ]:= << xAct`xCoba`
```

```
-----------------------------------------------------------
Package xAct`xCoba`   version 0.8.5, {2020, 2, 16}
CopyRight (C) 2005-2020, David Yllanes
  and Jose M. Martin-Garcia, under the General Public License.
-----------------------------------------------------------
These packages come with ABSOLUTELY NO WARRANTY; for details type
  Disclaimer[]. This is free software, and you are welcome to redistribute
  it under certain conditions. See the General Public License for details.
-----------------------------------------------------------
```

*In[◦]:=* `$DefInfoQ = False;`
`$PrePrint = ScreenDollarIndices;`
`$CVSimplify = Simplify;`

*In[◦]:=* `DefChart[cb, M4, {0, 1, 2, 3}, {t[], r[], θ[], ϕ[]}]`

*In[◦]:=* `cb /: CIndexForm[0, cb] := "t";`
`cb /: CIndexForm[1, cb] := "r";`
`cb /: CIndexForm[2, cb] := "θ";`
`cb /: CIndexForm[3, cb] := "ϕ";`

*In[◦]:=* `DefConstantSymbol[M]`
`DefConstantSymbol[a]`
`DefScalarFunction[f]`
`f[r[]] := 1 - 2 M/r[] + r[]^2/a^2;`

*In[◦]:=* `MatrixForm[met = DiagonalMatrix[{-f[r[]], f[r[]]^(-1), r[]^2, r[]^2 Sin[θ[]]^2}]]`

*Out[◦]//MatrixForm=*

$$\begin{pmatrix} -1 + \frac{2M}{r} - \frac{r^2}{a^2} & 0 & 0 & 0 \\ 0 & \frac{1}{1 - \frac{2M}{r} + \frac{r^2}{a^2}} & 0 & 0 \\ 0 & 0 & r^2 & 0 \\ 0 & 0 & 0 & r^2 \sin[\theta]^2 \end{pmatrix}$$

*In[◦]:=*

*In[◦]:=* `MetricInBasis[metric, -cb, met] // TableForm`

Added independent rule $g_{tt} \to -1 + \frac{2M}{r} - \frac{r^2}{a^2}$ for tensor metric

Added independent rule $g_{tr} \to 0$ for tensor metric

Added independent rule $g_{t\theta} \to 0$ for tensor metric

Added independent rule $g_{t\phi} \to 0$ for tensor metric

Added dependent rule $g_{rt} \to g_{tr}$ for tensor metric

Added independent rule $g_{rr} \to \frac{1}{1 - \frac{2M}{r} + \frac{r^2}{a^2}}$ for tensor metric

Added independent rule $g_{r\theta} \to 0$ for tensor metric

Added independent rule $g_{r\phi} \to 0$ for tensor metric

Added dependent rule $g_{\theta t} \to g_{t\theta}$ for tensor metric

Added dependent rule $g_{\theta r} \to g_{r\theta}$ for tensor metric

Added independent rule $g_{\theta\theta} \to r^2$ for tensor metric

Added independent rule $g_{\theta\phi} \to 0$ for tensor metric

Added dependent rule $g_{\phi t} \to g_{t\phi}$ for tensor metric

Added dependent rule $g_{\phi r} \to g_{r\phi}$ for tensor metric

Added dependent rule $g_{\phi\theta} \to g_{\theta\phi}$ for tensor metric

Added independent rule $g_{\phi\phi} \to r^2 \sin[\theta]^2$ for tensor metric

*Out[ ]//TableForm=*

| | | | |
|---|---|---|---|
| $g_{tt} \to -1 + \frac{2M}{r} - \frac{r^2}{a^2}$ | $g_{tr} \to 0$ | $g_{t\theta} \to 0$ | $g_{t\phi} \to 0$ |
| $g_{rt} \to 0$ | $g_{rr} \to \frac{1}{1 - \frac{2M}{r} + \frac{r^2}{a^2}}$ | $g_{r\theta} \to 0$ | $g_{r\phi} \to 0$ |
| $g_{\theta t} \to 0$ | $g_{\theta r} \to 0$ | $g_{\theta\theta} \to r^2$ | $g_{\theta\phi} \to 0$ |
| $g_{\phi t} \to 0$ | $g_{\phi r} \to 0$ | $g_{\phi\theta} \to 0$ | $g_{\phi\phi} \to r^2 \sin[\theta]^2$ |

*In[ ]:=* **TensorValues@metric**

*Out[ ]=* FoldedRule$\Big[ \{g_{rt} \to g_{tr},\ g_{\theta t} \to g_{t\theta},\ g_{\theta r} \to g_{r\theta},\ g_{\phi t} \to g_{t\phi},\ g_{\phi r} \to g_{r\phi},\ g_{\phi\theta} \to g_{\theta\phi}\},$
$\{g_{tt} \to -1 + \frac{2M}{r} - \frac{r^2}{a^2},\ g_{tr} \to 0,\ g_{t\theta} \to 0,\ g_{t\phi} \to 0,\ g_{rr} \to \frac{1}{1 - \frac{2M}{r} + \frac{r^2}{a^2}},$
$g_{r\theta} \to 0,\ g_{r\phi} \to 0,\ g_{\theta\theta} \to r^2,\ g_{\theta\phi} \to 0,\ g_{\phi\phi} \to r^2 \sin[\theta]^2 \}\Big]$

*In[ ]:=* **MetricCompute[metric, cb, "Weyl"[-1, -1, -1, -1]]**

```
In[○]:= Part[TensorValues@Ricci[CD][-α, -β]] // TableForm
```

Out[○]//TableForm=

$$\text{FoldedRule}\Big[\big\{R[\nabla]_{rt} \to R[\nabla]_{tr},\ R[\nabla]_{\theta t} \to R[\nabla]_{t\theta},\ R[\nabla]_{\theta r} \to R[\nabla]_{r\theta},$$
$$R[\nabla]_{\phi t} \to R[\nabla]_{t\phi},\ R[\nabla]_{\phi r} \to R[\nabla]_{r\phi},\ R[\nabla]_{\phi \theta} \to R[\nabla]_{\theta\phi}\big\},$$
$$\Big\{R[\nabla]_{tt} \to \frac{3\,(r^3 + a^2\,(-2M+r))}{a^4\,r},\ R[\nabla]_{tr} \to 0,\ R[\nabla]_{t\theta} \to 0,\ R[\nabla]_{t\phi} \to 0,$$
$$R[\nabla]_{rr} \to -\frac{3\,r}{r^3 + a^2\,(-2M+r)},\ R[\nabla]_{r\theta} \to 0,\ R[\nabla]_{r\phi} \to 0,$$
$$R[\nabla]_{\theta\theta} \to -\frac{3\,r^2}{a^2},\ R[\nabla]_{\theta\phi} \to 0,\ R[\nabla]_{\phi\phi} \to -\frac{3\,r^2\,\sin[\theta]^2}{a^2}\Big\}\Big]$$

```
In[○]:= Christoffel[CD, PDcb][α, -β, -γ]
```

Out[○]= $\Gamma[\nabla, \mathcal{D}]^{\alpha}{}_{\beta\gamma}$

```
In[○]:= Part[TensorValues@ChristoffelCDPDcb, 2] // TableForm
```

Out[○]//TableForm=

$\Gamma[\nabla,\mathcal{D}]^{t}{}_{tt} \to 0$

$\Gamma[\nabla,\mathcal{D}]^{t}{}_{tr} \to \frac{a^2 M + r^3}{r^4 + a^2 r\,(-2M+r)}$

$\Gamma[\nabla,\mathcal{D}]^{t}{}_{t\theta} \to 0$

$\Gamma[\nabla,\mathcal{D}]^{t}{}_{t\phi} \to 0$

$\Gamma[\nabla,\mathcal{D}]^{t}{}_{rr} \to 0$

$\Gamma[\nabla,\mathcal{D}]^{t}{}_{r\theta} \to 0$

$\Gamma[\nabla,\mathcal{D}]^{t}{}_{r\phi} \to 0$

$\Gamma[\nabla,\mathcal{D}]^{t}{}_{\theta\theta} \to 0$

$\Gamma[\nabla,\mathcal{D}]^{t}{}_{\theta\phi} \to 0$

$\Gamma[\nabla,\mathcal{D}]^{t}{}_{\phi\phi} \to 0$

$\Gamma[\nabla,\mathcal{D}]^{r}{}_{tt} \to \left(\frac{M}{r^2} + \frac{r}{a^2}\right)\left(1 - \frac{2M}{r} + \frac{r^2}{a^2}\right)$

$\Gamma[\nabla,\mathcal{D}]^{r}{}_{tr} \to 0$

$\Gamma[\nabla,\mathcal{D}]^{r}{}_{t\theta} \to 0$

$\Gamma[\nabla,\mathcal{D}]^{r}{}_{t\phi} \to 0$

$\Gamma[\nabla,\mathcal{D}]^{r}{}_{rr} \to -\frac{a^2 M + r^3}{r^4 + a^2 r\,(-2M+r)}$

$\Gamma[\nabla,\mathcal{D}]^{r}{}_{r\theta} \to 0$

$\Gamma[\nabla,\mathcal{D}]^{r}{}_{r\phi} \to 0$

$\Gamma[\nabla,\mathcal{D}]^{r}{}_{\theta\theta} \to 2M - r - \frac{r^3}{a^2}$

$\Gamma[\nabla,\mathcal{D}]^{r}{}_{\theta\phi} \to 0$

$\Gamma[\nabla,\mathcal{D}]^{r}{}_{\phi\phi} \to -\frac{(r^3 + a^2\,(-2M+r))\,\sin[\theta]^2}{a^2}$

$\Gamma[\nabla,\mathcal{D}]^{\theta}{}_{tt} \to 0$

$\Gamma[\nabla,\mathcal{D}]^{\theta}{}_{tr} \to 0$

$\Gamma[\nabla,\mathcal{D}]^{\theta}{}_{t\theta} \to 0$

$\Gamma[\nabla,\mathcal{D}]^{\theta}{}_{t\phi} \to 0$

$\Gamma[\nabla, \mathcal{D}]^{\theta}{}_{rr} \to 0$

$\Gamma[\nabla, \mathcal{D}]^{\theta}{}_{r\theta} \to \frac{1}{r}$

$\Gamma[\nabla, \mathcal{D}]^{\theta}{}_{r\phi} \to 0$

$\Gamma[\nabla, \mathcal{D}]^{\theta}{}_{\theta\theta} \to 0$

$\Gamma[\nabla, \mathcal{D}]^{\theta}{}_{\theta\phi} \to 0$

$\Gamma[\nabla, \mathcal{D}]^{\theta}{}_{\phi\phi} \to -\text{Cos}[\theta]\,\text{Sin}[\theta]$

$\Gamma[\nabla, \mathcal{D}]^{\phi}{}_{tt} \to 0$

$\Gamma[\nabla, \mathcal{D}]^{\phi}{}_{tr} \to 0$

$\Gamma[\nabla, \mathcal{D}]^{\phi}{}_{t\theta} \to 0$

$\Gamma[\nabla, \mathcal{D}]^{\phi}{}_{t\phi} \to 0$

$\Gamma[\nabla, \mathcal{D}]^{\phi}{}_{rr} \to 0$

$\Gamma[\nabla, \mathcal{D}]^{\phi}{}_{r\theta} \to 0$

$\Gamma[\nabla, \mathcal{D}]^{\phi}{}_{r\phi} \to \frac{1}{r}$

$\Gamma[\nabla, \mathcal{D}]^{\phi}{}_{\theta\theta} \to 0$

$\Gamma[\nabla, \mathcal{D}]^{\phi}{}_{\theta\phi} \to \text{Cot}[\theta]$

$\Gamma[\nabla, \mathcal{D}]^{\phi}{}_{\phi\phi} \to 0$

```mathematica
In[1]:= (* SAdS4 Metric *)
coords = {t, r, θ, ϕ};
n = Length[coords];
tt = -(1 - (2 M / r) + (r^2 / a^2));
rr = 1 / (1 - (2 M / r) + (r^2 / a^2));
θθ = r^2;
φφ = r^2 Sin[θ]^2;
gdd = {{tt, 0, 0, 0}, {0, rr, 0, 0}, {0, 0, θθ, 0}, {0, 0, 0, φφ}};
Simplify[gdd] // MatrixForm
guu = Simplify[Inverse[gdd]];
guu // MatrixForm
```

Out[8]//MatrixForm=
$$\begin{pmatrix} -1 + \frac{2M}{r} - \frac{r^2}{a^2} & 0 & 0 & 0 \\ 0 & \frac{1}{1 - \frac{2M}{r} + \frac{r^2}{a^2}} & 0 & 0 \\ 0 & 0 & r^2 & 0 \\ 0 & 0 & 0 & r^2 \text{Sin}[\theta]^2 \end{pmatrix}$$

Out[10]//MatrixForm=
$$\begin{pmatrix} \frac{1}{-1 + \frac{2M}{r} - \frac{r^2}{a^2}} & 0 & 0 & 0 \\ 0 & 1 - \frac{2M}{r} + \frac{r^2}{a^2} & 0 & 0 \\ 0 & 0 & \frac{1}{r^2} & 0 \\ 0 & 0 & 0 & \frac{\text{Csc}[\theta]^2}{r^2} \end{pmatrix}$$

```
In[11]:= (* Christoffel Symbols for SAdS4. Compare with xTensor calculations above. *)
christoffel := christoffel = Simplify[Table[(1 / 2) *
    Sum[(guu[[i, s]]) * (D[gdd[[s, j]], coords[[k]]] + D[gdd[[s, k]], coords[[j]]] -
        D[gdd[[j, k]], coords[[s]]]), {s, 1, n}], {i, 1, n}, {j, 1, n}, {k, 1, n}]]
listchristoffel := Table[If[UnsameQ[christoffel[[i, j, k]], 0],
    {ToString[Γ[i, j, k]], christoffel[[i, j, k]]}], {i, 1, n}, {j, 1, n}, {k, 1, j}]
TableForm[Partition[DeleteCases[Flatten[listchristoffel], Null], 2],
 TableSpacing → {2, 2}]
geodesic := geodesic = Simplify[
   Table[-Sum[christoffel[[i, j, k]] coords[[j]]' coords[[k]]', {j, 1, n}, {k, 1, n}],
     {i, 1, n}]]
listgeodesic := Table[{"d/dτ" ToString[coords[[i]]'], "=", geodesic[[i]]},
  {i, 1, n}]
TableForm[Partition[DeleteCases[Flatten[listgeodesic], Null], 2],
 TableSpacing → {2, 2}]
```

Out[13]//TableForm=

| $\Gamma[1, 2, 1]$ | $\frac{a^2 M + r^3}{r^4 + a^2 r (-2M+r)}$ |
| --- | --- |
| $\Gamma[2, 1, 1]$ | $\frac{(a^2 M + r^3)(r^3 + a^2(-2M+r))}{a^4 r^3}$ |
| $\Gamma[2, 2, 2]$ | $-\frac{a^2 M + r^3}{r^4 + a^2 r (-2M+r)}$ |
| $\Gamma[2, 3, 3]$ | $2M - r - \frac{r^3}{a^2}$ |
| $\Gamma[2, 4, 4]$ | $-\frac{(r^3 + a^2(-2M+r))\sin[\theta]^2}{a^2}$ |
| $\Gamma[3, 3, 2]$ | $\frac{1}{r}$ |
| $\Gamma[3, 4, 4]$ | $-\cos[\theta]\sin[\theta]$ |
| $\Gamma[4, 4, 2]$ | $\frac{1}{r}$ |
| $\Gamma[4, 4, 3]$ | $\cot[\theta]$ |

Out[16]//TableForm=

| d/dτ t' | = |
| --- | --- |
| $-\frac{2(a^2 M + r^3) r' t'}{r^4 + a^2 r (-2M+r)}$ | d/dτ r' |
| = | $\frac{(a^2 M + r^3)(r')^2}{r^4 + a^2 r (-2M+r)} + \frac{(r^3 + a^2(-2M+r))(-(a^2 M + r^3)(t')^2 + a^2 r^3((\theta')^2 + \sin[\theta]^2 (\phi')^2))}{a^4 r^3}$ |
| d/dτ θ' | = |
| $-\frac{2 r' \theta'}{r} + \cos[\theta]\sin[\theta](\phi')^2$ | d/dτ φ' |
| = | $-\frac{2(r' + r \cot[\theta] \theta')\phi'}{r}$ |

```
In[17]:= (* For taking Covariant Derivative of Type-(0, 1) Tensors *)
CovDer01[u_] := Block[{uα}, uα = u;
  Table[(D[uα[[i]], coords[[j]]] - Sum[uα[[k]] * christoffel[[k, i, j]], {k, 1, n}]),
    {i, 1, n}, {j, 1, n}]]
```

```
In[18]:= (* r-component of contravariant 4-Velocity, u. Refer to Eq.(6) in paper. *)
        uru = Sqrt[2 M / r - r^2 / a^2 - L^2 / (r^2 Sin[θ]^2) (1 - 2 M / r + r^2 / a^2)];
        urd = rr * uru; (* Covariant r-component *)
        ud = {-1, urd, 0, L} (* Covariant 4-Velocity *)
```

$$\text{Out[20]= } \left\{-1,\ \frac{\sqrt{\frac{2M}{r} - \frac{r^2}{a^2} - \frac{L^2\left(1 - \frac{2M}{r} + \frac{r^2}{a^2}\right)\operatorname{Csc}[\theta]^2}{r^2}}}{1 - \frac{2M}{r} + \frac{r^2}{a^2}},\ 0,\ L\right\}$$

```
In[21]:= (* Covariant form of Expansion Tensor, B^a_b *)
        B = Simplify[CovDer01[ud]];
        (* B//MatrixForm *)

In[22]:= (* Symmetric part of B *)
        sym = Table[(1 / 2) * (B[[i, j]] + B[[j, i]]), {i, 1, n}, {j, 1, n}];
        (* sym//MatrixForm *)

In[23]:= (* Covariant Projection tensor *)
        hdd = Simplify[Table[gdd[[i, j]] + Outer[Times, ud, ud][[i, j]], {i, 1, n}, {j, 1, n}]];
        hdd // MatrixForm
```

$$\text{Out[24]//MatrixForm= } \begin{pmatrix} \frac{2M}{r} - \frac{r^2}{a^2} & -\frac{a^2 r \sqrt{\frac{2Mr - \frac{r^4}{a^2} + L^2\left(-1 + \frac{2M}{r} - \frac{r^2}{a^2}\right)\operatorname{Csc}[\theta]^2}{r^2}}}{r^3 + a^2(-2M + r)} & 0 & -L \\ -\frac{a^2 r \sqrt{\frac{2Mr - \frac{r^4}{a^2} + L^2\left(-1 + \frac{2M}{r} - \frac{r^2}{a^2}\right)\operatorname{Csc}[\theta]^2}{r^2}}}{r^3 + a^2(-2M + r)} & \frac{a^4 r^3 + a^2 L^2\left(a^2(2M - r) - r^3\right)\operatorname{Csc}[\theta]^2}{r\left(r^3 + a^2(-2M + r)\right)^2} & 0 & \frac{a^2 L r \sqrt{\frac{2Mr - \frac{r^4}{a^2} + L^2\left(-1 + \frac{2M}{r} - \frac{r^2}{a^2}\right)\operatorname{Csc}[\theta]^2}{r^2}}}{r^3 + a^2(-2M + r)} \\ 0 & 0 & r^2 & 0 \\ -L & \frac{a^2 L r \sqrt{\frac{2Mr - \frac{r^4}{a^2} + L^2\left(-1 + \frac{2M}{r} - \frac{r^2}{a^2}\right)\operatorname{Csc}[\theta]^2}{r^2}}}{r^3 + a^2(-2M + r)} & 0 & L^2 + r^2 \operatorname{Sin}[\theta]^2 \end{pmatrix}$$

```
In[25]:= (* Setting coordinate θ to π/2, as we are working in the equatorial plane. *)
        θ = π / 2;

In[26]:= (* Expansion Scalar - Eq.(7) *)
        Θ_T = Simplify[Tr[guu.B]]
```

$$\text{Out[26]= } \frac{-2L^2 r^3 - 3r^5 + a^2\left(L^2(M - r) + 3Mr^2\right)}{a^2 r^4 \sqrt{\frac{L^2(2M - r)}{r^3} + \frac{2M}{r} - \frac{L^2 + r^2}{a^2}}}$$

```
In[30]:= (* Rotation tensor and its quadratic invariant *)
        ωdd = Simplify[(1 / 2) Table[(B[[i, j]] - B[[j, i]]), {i, 1, n}, {j, 1, n}]];
        ωuu = Simplify[guu.ωdd.guu];
        ω = Tr[Simplify[ωuu.ωdd]]
```

Out[32]= 0

```
In[33]:= (* Shear tensor and its quadratic invariant *)
        σdd = Simplify[- (1 / 3) * Θ_T * hdd + sym];
        σuu = Simplify[guu.σdd. guu];
        σ = Tr[Simplify[σuu.σdd]]
```

$$\text{Out[35]=}\quad \frac{\left(L^2 r^3 + a^2 \left(-3 M r^2 + L^2 (-5 M + 2 r)\right)\right)^2}{-9 a^2 r^8 \left(L^2 + r^2\right) + 9 a^4 \left(L^2 (2 M - r) r^5 + 2 M r^7\right)} +$$

$$\left(4 L^4 r^6 \left(L^2 + r^2\right) + a^2 \left(8 L^4 (M - r) r^5 + 3 L^2 (4 M - 3 r) r^7 + 4 L^6 r^3 (-M + r)\right) + \right.$$
$$\left. a^4 \left(L^6 (M - r)^2 - 5 L^4 (M - r)^2 r^2 + 9 M^2 r^6 + 3 L^2 M r^4 (M + 2 r)\right)\right) \big/$$
$$\left(9 a^2 r^7 \left(-r^3 \left(L^2 + r^2\right) + a^2 \left(L^2 (2 M - r) + 2 M r^2\right)\right)\right) +$$
$$\left(L^4 r^9 + a^2 \left(L^4 (6 M - 5 r) r^6 + 3 L^2 (4 M - 3 r) r^8\right) + \right.$$
$$\left. a^4 r^4 \left(L^4 (8 M - 5 r) + 36 M^2 r^3 + 6 L^2 M r (4 M + r)\right) - \right.$$
$$\left. a^6 M \left(3 L^2 M (32 M - 17 r) r^2 + 72 M^2 r^4 + L^4 \left(32 M^2 - 31 M r + 8 r^2\right)\right)\right) \big/$$
$$\left(9 a^2 r^5 \left(r^3 + a^2 (-2 M + r)\right) \left(-r^3 \left(L^2 + r^2\right) + a^2 \left(L^2 (2 M - r) + 2 M r^2\right)\right)\right) +$$
$$\left(-4 L^6 r^9 + 4 a^2 \left(L^6 (3 M - 2 r) r^6 + L^4 r^8 (-3 M + 4 r)\right) + \right.$$
$$\left. a^6 \left(L^6 (M - r)^2 (2 M - r) - L^4 (12 M - 7 r) (M - r)^2 r^2 + 36 M^2 r^7 + 6 L^2 M r^4 \left(3 M^2 + 8 M r - 5 r^2\right)\right) - \right.$$
$$\left. a^4 \left(3 L^2 r^7 \left(3 M^2 + 2 M r + 3 r^2\right) + L^6 r^3 \left(9 M^2 - 14 M r + 5 r^2\right) - 2 L^4 r^5 \left(15 M^2 - 26 M r + 11 r^2\right)\right)\right) \big/$$
$$\left(9 a^2 r^7 \left(r^3 + a^2 (-2 M + r)\right) \left(-r^3 \left(L^2 + r^2\right) + a^2 \left(L^2 (2 M - r) + 2 M r^2\right)\right)\right)$$

```
In[36]:= (* Covariant Ricci Tensor Terms, calculated using xTensor above *)
        Rtt = 3 (r^3 + a^2 (-2 M + r)) / (a^4 r);
        Rrr = -3 r / (r^3 + a^2 (-2 M + r));
        Rθθ = -3 r^2 / a^2;
        Rϕϕ = -3 r^2 Sin[θ]^2 / a^2;
        Rdd = Simplify[{{Rtt, 0, 0, 0}, {0, Rrr, 0, 0}, {0, 0, Rθθ, 0}, {0, 0, 0, Rϕϕ}}];
        Rdd // MatrixForm
```

$$\text{Out[41]//MatrixForm=}\quad \begin{pmatrix} \frac{3 \left(r^3 + a^2 (-2 M + r)\right)}{a^4 r} & 0 & 0 & 0 \\ 0 & -\frac{3 r}{r^3 + a^2 (-2 M + r)} & 0 & 0 \\ 0 & 0 & -\frac{3 r^2}{a^2} & 0 \\ 0 & 0 & 0 & -\frac{3 r^2}{a^2} \end{pmatrix}$$

```
In[42]:= (* Contravariant 4-Velocity *)
        uu = Simplify[guu.ud];
        uu // MatrixForm
```

$$\text{Out[43]//MatrixForm=}\quad \begin{pmatrix} \frac{a^2 r}{r^3 + a^2 (-2 M + r)} \\ \sqrt{\frac{2 M r - \frac{r^4}{a^2} + L^2 \left(-1 + \frac{2 M}{r} - \frac{r^2}{a^2}\right)}{r^2}} \\ 0 \\ \frac{L}{r^2} \end{pmatrix}$$

In[44]:= `(* Calculating Raychaudhuri scalar *)`
`rayscalar = Simplify[Tr[Rdd.Outer[Times, uu, uu]]]`

Out[44]= $\dfrac{3}{a^2}$

In[45]:= `congruence = Simplify[- (1 / 3) (`$\Theta_T$`^2) - σ - rayscalar]`

Out[45]= $\left(-L^2 r^6 \left(2 L^2 + r^2\right) + a^2 \left(2 L^4 (M-r) r^3 + L^2 r^5 (-2M + 3r)\right) - a^4 \left(2 L^2 M (5M - 3r) r^2 + 9 M^2 r^4 + L^4 \left(5 M^2 - 4 M r + r^2\right)\right)\right) / \left(-a^2 r^8 \left(L^2 + r^2\right) + a^4 \left(L^2 (2M - r) r^5 + 2 M r^7\right)\right)$

In[46]:= `(* Special case. Result matches Eq.(7), with L`$_T$` = 0. *)`
`L = 0;`
`Simplify[congruence]`

Out[47]= $\dfrac{9 a^2 M^2}{-2 a^2 M r^3 + r^6}$

# APPENDIX B - CODE FOR RAYCHAUDHARI EQUATION OF NULLLIKE GEODESICS

```mathematica
(* Using definitions of gdd, guu, christoffels and CovDer01 from Appendix A *)
```

In[ ]:= `(* Covariant 4-Velocity *)`
```mathematica
kd = {-1, Sqrt[1 - (L^2 (r^3 + a^2 (-2 M + r)) Csc[θ]^2) / (a^2 r^3)] /
    (1 - 2 M / r + r^2 / a^2), 0, L }
```

Out[ ]= $\left\{-1, \dfrac{\sqrt{1 - \dfrac{L^2 \left(r^3 + a^2 (-2M+r)\right) \text{Csc}[\theta]^2}{a^2 r^3}}}{1 - \dfrac{2M}{r} + \dfrac{r^2}{a^2}}, 0, L\right\}$

In[ ]:= `(* Contravariant form of auxilliary null vector field, n *)`
```mathematica
nu = { (-a^2 r^3 Sin[θ]^2 + Sqrt[a^2 L^2 r^3 (r^3 + a^2 (-2 M + r)) Sin[θ]^2]) /
    (L^2 (r^3 + a^2 (-2 M + r)) - a^2 r^3 Sin[θ]^2), 0, 0,
    (L^2 r^3 + a^2 L^2 (-2 M + r) - Sqrt[a^2 L^2 r^3 (r^3 + a^2 (-2 M + r)) Sin[θ]^2]) /
    (L^3 (a^2 (2 M - r) - r^3) + a^2 L r^3 Sin[θ]^2) };
nd = Simplify[gdd.nu]; (* Covariant n *)
nd // MatrixForm
```

Out[ ]//MatrixForm=
$$\begin{pmatrix} \dfrac{\left(-1+\dfrac{2M}{r}-\dfrac{r^2}{a^2}\right)\left(-a^2 r^3 \sin[\theta]^2 + \sqrt{a^2 L^2 r^3 \left(r^3+a^2(-2M+r)\right)\sin[\theta]^2}\right)}{L^2\left(r^3+a^2(-2M+r)\right)-a^2 r^3 \sin[\theta]^2} \\ 0 \\ 0 \\ -\dfrac{r^2 \sin[\theta]^2 \left(L^2 r^3 + a^2 L^2(-2M+r) - \sqrt{a^2 L^2 r^3\left(r^3+a^2(-2M+r)\right)\sin[\theta]^2}\right)}{L^3\left(r^3+a^2(-2M+r)\right)-a^2 L r^3 \sin[\theta]^2} \end{pmatrix}$$

In[ ]:= `(* k & n satisfy k_a n^a = -1 and n_a n^a = 0 *)`
```mathematica
Simplify[kd.nu]
Simplify[nd.nu]
```

Out[ ]= $-1$

Out[ ]= $0$

In[ ]:= `(* Covariant form of Expansion Tensor, B^a_b *)`
```mathematica
B = Simplify[CovDer01[kd]];
(* B//MatrixForm *)
```

In[ ]:= `(* Covariant Projection tensor *)`
```mathematica
hdd = Simplify[Table[gdd[[i, j]] + Outer[Times, kd, nd][[i, j]] +
    Outer[Times, kd, nd][[j, i]], {i, 1, n}, {j, 1, n}]];
(* hdd//MatrixForm *)
```

In[ ]:= `(* Mixed form of Projection tensor *)`
```mathematica
hud = Simplify[guu.hdd];
BTil = Simplify[Table[Sum[hud[[b, d]] * B[[b, c]] * hud[[c, e]], {b, 1, n}, {c, 1, n}],
    {d, 1, n}, {e, 1, n}]];
(* BTil//MatrixForm *)
```

```mathematica
In[•]:= (* Checking Θ_N = g^ab BTil_ab = g^ab B_ab *)
    Simplify[Tr[guu.BTil]]
    Simplify[Tr[guu.B]]
```

$$\text{Out[•]= } \frac{2 a^2 r^3 + L^2 \left(a^2 (M - r) - 2 r^3\right) \csc[\theta]^2}{a^2 r^4 \sqrt{1 - \frac{L^2 \left(r^3 + a^2 (-2 M + r)\right) \csc[\theta]^2}{a^2 r^3}}}$$

$$\text{Out[•]= } \frac{2 a^2 r^3 + L^2 \left(a^2 (M - r) - 2 r^3\right) \csc[\theta]^2}{a^2 r^4 \sqrt{1 - \frac{L^2 \left(r^3 + a^2 (-2 M + r)\right) \csc[\theta]^2}{a^2 r^3}}}$$

```mathematica
(* Rotation tensor and its quadratic invariant *)
ωdd = Simplify[(1/2) Table[(BTil[[i, j]] - BTil[[j, i]]), {i, 1, n}, {j, 1, n}]];
ωuu = Simplify[guu.ωdd.guu];
ω = Tr[Simplify[ωuu.ωdd]]
```

$$\text{Out[•]= } \frac{L^4 \left(r^3 + a^2 (-2M + r)\right) \cot[\theta]^2 \csc[\theta]^2}{4 L^2 r^7 - 2 a^2 r^4 \left(L^2 (4M - 2r) + r^3\right) + 2 a^2 r^7 \cos[2\theta]} -$$

$$\frac{L^4 \left(r^3 + a^2 (-2M + r)\right) \cot[\theta]^2 \csc[\theta]^4}{4 r^4 \left(a^2 r^3 + L^2 \left(a^2 (2M - r) - r^3\right) \csc[\theta]^2\right)} -$$

$$\left(L^2 \cot[\theta]^2 \csc[\theta]^6 \left(L^2 r^3 + a^2 L^2 (-2M + r) - \sqrt{a^2 L^2 r^3 \left(r^3 + a^2 (-2M + r)\right) \sin[\theta]^2}\right)^2\right) \Big/$$

$$\left(4 r^4 \left(a^2 r^3 + L^2 \left(a^2 (2M - r) - r^3\right) \csc[\theta]^2\right)^2\right) + \left(L^4 \left(r^3 + a^2 (-2M + r)\right) \cot[\theta]^2\right.$$

$$\left.\csc[\theta]^8 \left(-a^2 r^3 + a^2 r^3 \cos[2\theta] + 2 \sqrt{a^2 L^2 r^3 \left(r^3 + a^2 (-2M + r)\right) \sin[\theta]^2}\right)^2\right) \Big/$$

$$\left(16 r^7 \left(a^3 r^3 + a L^2 \left(a^2 (2M - r) - r^3\right) \csc[\theta]^2\right)^2\right)$$

```mathematica
In[•]:= (* Symmetric part of BTil *)
    sym = Simplify[(1/2) Table[(BTil[[i, j]] + BTil[[j, i]]), {i, 1, n}, {j, 1, n}]];
    (* sym//MatrixForm *)

In[•]:= (*Setting coordinate θ to π/2,as we are working in the equatorial plane.*)
    θ = π/2;

In[•]:= (* Expansion Scalar - Eq.(13) *)
    Θ_N = Simplify[Tr[guu.B]]
```

$$\text{Out[•]= } \frac{-2 L^2 r^3 + a^2 \left(L^2 (M - r) + 2 r^3\right)}{a^2 \sqrt{1 - \frac{L^2}{a^2} + \frac{L^2 (2M - r)}{r^3}} \, r^4}$$

```
In[ ]:= (* Shear tensor and its quadratic invariant *)
    σdd = Simplify[-(1/2)*Θ_N*hdd + sym];
    σuu = Simplify[guu.σdd.guu];
    σ = Tr[Simplify[σuu.σdd]]
```

$$\text{Out[ ]=}\quad \frac{L^4 (-3M+r)^2}{4\left(1-\frac{L^2}{a^2}+\frac{L^2(2M-r)}{r^3}\right)r^8} + \frac{a^2 L^4 (-3M+r)^2}{-4 L^2 r^8 + 4 a^2 \left(L^2 (2M-r) r^5 + r^8\right)}$$

```
In[ ]:= (* Covariant Ricci Tensor Terms, calculated using xTensor *)
    Rtt = 3 (r^3 + a^2 (-2 M + r)) / (a^4 r);
    Rrr = -3 r / (r^3 + a^2 (-2 M + r));
    Rθθ = -3 r^2 / a^2;
    Rϕϕ = -3 r^2 Sin[θ]^2 / a^2;
    Rdd = Simplify[{{Rtt, 0, 0, 0}, {0, Rrr, 0, 0}, {0, 0, Rθθ, 0}, {0, 0, 0, Rϕϕ}}];
    Rdd // MatrixForm
```

Out[ ]//MatrixForm=

$$\begin{pmatrix} \frac{3(r^3+a^2(-2M+r))}{a^4 r} & 0 & 0 & 0 \\ 0 & -\frac{3r}{r^3+a^2(-2M+r)} & 0 & 0 \\ 0 & 0 & -\frac{3r^2}{a^2} & 0 \\ 0 & 0 & 0 & -\frac{3r^2}{a^2} \end{pmatrix}$$

```
In[ ]:= (* Contravariant 4-Velocity, k *)
    ku = Simplify[guu.kd];
    ku // MatrixForm
```

Out[ ]//MatrixForm=

$$\begin{pmatrix} \frac{a^2 r}{r^3+a^2(-2M+r)} \\ \sqrt{1-\frac{L^2}{a^2}+\frac{L^2(2M-r)}{r^3}} \\ 0 \\ \frac{L}{r^2} \end{pmatrix}$$

```
In[ ]:= (* Calculating Raychaudhuri scalar *)
    rayscalar = Simplify[Tr[Rdd.Outer[Times, ku, ku]]]
```

Out[ ]= 0

```
In[ ]:= congruence = Simplify[-(1/2) (Θ_N^2) - σ + ω - rayscalar]
```

$$\text{Out[ ]=}\quad \left(-2 L^4 r^6 + 2 a^2 \left(L^4 (M-r) r^3 + 2 L^2 r^6\right) - a^4 \left(2 L^2 (M-r) r^3 + 2 r^6 + L^4 \left(5 M^2 - 4 M r + r^2\right)\right)\right) / \left(-a^2 L^2 r^8 + a^4 \left(L^2 (2M-r) r^5 + r^8\right)\right)$$



\end{document}